\documentclass{aa}

\usepackage{txfonts}
\usepackage{color}
\usepackage{natbib}
\bibpunct{(}{)}{;}{a}{}{,} 
\usepackage{graphicx}
\usepackage{longtable, lscape}

\title{The projection factor of $\delta$~Cephei}

\subtitle{A calibration of the Baade-Wesselink method using the CHARA
  Array}

\titlerunning{}
\author{ 
  Antoine M\'erand\inst{1}\thanks{To whom correspondence should be
    addressed, \texttt{antoine.merand@obspm.fr}} \and
  Pierre Kervella\inst{1} \and  
  Vincent Coud\'e du Foresto\inst{1} \and
  Stephen~T.  Ridgway\inst{1,2,3} \and  
  Jason~P. Aufdenberg\inst{2} \and
  Theo~A. ten  Brummelaar\inst{3} \and  
  David~H. Berger\inst{3} \and  
  Judit Sturmann\inst{3} \and
  Laszlo Sturmann\inst{3} \and  
  Nils~H. Turner\inst{3} \and  
  Harold~A. McAlister\inst{3}}
\authorrunning{M\'erand et al.}
\institute{LESIA, UMR8109, Observatoire de Paris, 
  5, place Jules Janssen, 92195 Meudon (France) \and
  National Optical Astronomical Observatory
  950 North Cherry Avenue, Tucson, AZ 85719 (USA)\and
  Center for High Angular Resolution Astronomy,
  Georgia State University, P.O. Box 3965, Atlanta, Georgia 30302-3965 (USA)}

\date{Received --- / Accepted ---} 

\abstract{ Cepheids play a key role in astronomy as standard candles
for measuring intergalactic distances. Their distance is usually
inferred from the Period-Luminosity relationship, calibrated using the
semi-empirical Baade-Wesselink method. Using this method, the distance
is known to a multiplicative factor, called the projection
factor. Presently, this factor is computed using numerical models - it
has hitherto never been measured directly. Based on our new
interferometric measurements obtained with the CHARA Array and the
already published parallax, we present a geometrical measurement of
the projection factor of a Cepheid, \object{$\delta$~Cep}. The value
we determined, $p=1.27\pm0.06$, confirms the generally adopted value
of $p=1.36$ within 1.5 sigmas. Our value is in line with recent
theoretical predictions of \citet{Nardetto2004}.


}
				   
\begin{document} 
\maketitle

\section{Introduction}

Cepheid stars are commonly used as cosmological distance indicators,
thanks to their well-established Period-Luminosity law (P-L). This
remarkable property has turned these supergiant stars into primary
standard candles for extragalactic distance estimations.  With
intrinsic brightnesses of up to 100,000 times that of the Sun,
Cepheids are easily distinguished in distant galaxies (up to about 30
Mpc distant). As such, they are used to calibrate the secondary
distance indicators (supernovae, etc...) that are used to estimate
even larger cosmological distances.  For instance, the {\it Hubble Key
Project} to measure the Hubble constant $H_0$ \citep{Freedman2001} is
based on the assumption of a distance to the LMC that was established
primarily using Cepheids.  Located at the very base of the
cosmological distance ladder, a bias on the calibration of the Cepheid
P-L relation would impact our whole perception of the scale of the
Universe.

\subsection{Period-Luminosity calibration} 
The P-L relation takes the form $\log L = \alpha \log P + \beta$,
where $L$ is the (absolute) luminosity, $P$ the period, $\alpha$ the
slope, and $\beta$ the zero point.  The determination of $\alpha$ is
straightforward: one can consider a large number of Cepheids in the
LMC, located at a common distance from us.  Calibrating the zero-point
$\beta$ is a much more challenging task, as it requires an independent
distance measurement to a number of Cepheids.  Ideally, one should
measure directly their geometrical parallaxes, in order to obtain
their absolute luminosity.  Knowing their variation period, $\beta$
would then come out easily. However, Cepheids are rare stars: only a
few of them are located in the solar neighborhood, and these nearby
stars are generally too far away for precise parallax measurements,
with the exception of $\delta$~Cep.

\subsection{The Baade-Wesselink method} 
The most commonly used alternative to measure the distance to a
pulsating star is the Baade-Wesselink (BW) method.  Developed in the
first part of the 20th century \citep{Baade1926, Wesselink1946}, it
utilizes the pulsational velocity $V_\mathrm{puls.}$ of the surface of
the star and its angular size.  Integrating the pulsational velocity
curve provides an estimation of the linear radius variation
over the pulsation. Comparing the {\it linear} and {\it angular}
amplitudes of the Cepheid pulsation gives directly its distance.  The
most recent implementation \citep{Kervella2004} of the BW method makes
use of long-baseline interferometry to measure directly the angular
size of the star.

Unfortunately, spectroscopy measures the apparent radial velocity
$V_\mathrm{rad.}$, i.e. the Doppler shift of absorption lines in the
stellar atmosphere, projected along the line of sight and integrated
over the stellar disk. This is where $p$, a projection factor, has to
be introduced, which is defined as
$p=V_\mathrm{puls.}/V_\mathrm{rad.}$.  The general BW method can be
summarized in the relation:

\begin{equation}
\theta(T)-\theta(0) = - 2\,\frac{p}{d}\int_0^T
(V_\mathrm{rad.}(t)-V_\gamma)\,dt
\label{bw}
\end{equation}
where $d$ is the distance, $p$ the projection factor, $\theta$ the
angular diameter and $V_\gamma$ the systematic radial velocity.  There
are in fact many contributors to the $p$-factor. The main ones are the
sphericity of the star (purely geometrical) and its limb darkening
(due to the stellar atmosphere structure). A careful theoretical
calculation of $p$ requires modeling dynamically the formation of the
absorption line in the pulsating atmosphere of the Cepheid
\citep{Parsons1972, Sabbey1995, Nardetto2004}.

Until now, distance measurements to Cepheids used a $p$-factor value
estimated from numerical models.  Looking closely at Eq.~\ref{bw}, it
is clear that any uncertainty on the value of $p$ will create the same
relative uncertainty on the distance estimation, and
subsequently to the P-L relation calibration.  In other words, the
Cepheid distance scale relies implicitly on numerical models of these
stars. But how good are the models?  To answer this question, one
should confront their predictions to measurable quantities. Until now,
this comparison was impossible due to the difficulty to constrain the
two variables $\theta(T)$ and $d$ from observations, i.e. the angular
diameter and the distance.

Among classical Cepheids, {$\delta$~Cep} (\object{HR~8571},
\object{HD~213306}) is remarkable: it is not only the prototype of its
kind, but also the Cepheid with the most precise trigonometric
parallax currently available, obtained recently using the FGS
instrument aboard the {\it Hubble Space Telescope}
\citep{Benedict2002}. This direct measurement of the distance opens
the way to the direct measurement (with the smallest sensitivity to
stellar models) of the $p$ factor of $\delta$~Cep, provided that
high-precision angular diameters can be measured by interferometry.

\section{Application of the BW method to $\delta$~Cep.} 

To achieve this goal, interferometric observations were undertaken at
the CHARA Array \citep{tenBrummelaar2003, tenBrummelaar2005}, in the
infrared K' band ($1.95\,\mu\mathrm{m}\leq \lambda \leq
2.3\,\mu\mathrm{m} $) with the Fiber Linked Unit for Optical
Recombination \citep{Foresto2003} (FLUOR) using two East-West baselines
of the CHARA Array: E1-W1 and E2-W1, with baselines of 313 and 251 m
respectively. Observations took place during summer 2004 for E2-W1
(seven nights between JD~$2\ 453\ 216$ and JD~$2\ 453\ 233$) and Fall 2004
for E1-W1 (six consecutive nights, from JD~$2\ 453\ 280$ to JD~$2\
453\ 285$). The pulsation phase was computed using the following
period and reference epoch \citep{Moffett1985}: $P=5.366316$\,d, $T_0=
2\ 453\ 674.144$ (Julian date), the 0-phase being defined at maximum
light in the V band. The resulting phase coverage is very good for the
longest baseline (E1-W1), while data lack at minimum diameter for the
smaller one (E2-W1)

The FLUOR Data reduction software (DRS) \citep{Foresto1997}, was used
to extract the squared modulus of the coherence factor between the two
independent apertures. All calibrator stars were chosen in a catalogue
computed for this specific purpose \citep{Merand2005} (see
Table~\ref{calib1}). Calibrators chosen for this work are all K
giants, whereas $\delta$~Cep is a G0 supergiant. The spectral type
difference is properly taken into account in the reduction, even
though it has no significant influence on the final result.  The
interferometric transfer function of the instrument was estimated by
observing calibrators before and after each $\delta$~Cep data
point. The efficiency of CHARA/FLUOR was consistent between all
calibrators and stable over the night around 85\%. Data that share a
calibrator are affected by a common systematic error due to the
uncertainty of the {\it a priori} angular diameter of this
calibrator. In order to interpret our data properly, we used a
specific formalism \citep{Perrin2003} tailored to propagate these
correlations into the model fitting process.  Diameters are derived
from the visibility data points using a full model of the FLUOR
instrument including the spectral bandwidth effects
\citep{Kervella2003}.  The stellar center-to-limb darkening is
corrected using a model intensity profile taken from tabulated values
\citep{Claret2000} with parameters corresponding to $\delta$~Cep
($T_\mathrm{eff}=6000K$, $\log g=2.0$ and solar metallicity).  The
limb darkened (LD) angular diameter comes out 3\% larger than its
uniform disk (UD) counterpart.


\begin{table}
    \begin{center}
      \begin{tabular}{lccc}
        \hline  & S. Type & UD diam. (mas) & Baseline\\  
	\hline
	\object{HD 2952}    & K0III    & $0.938\pm 0.013$ & W1-E1\\ 
	\object{HD 138852}  & K0III-IV & $0.952\pm 0.012$ & W1-E1\\ 
	\object{HD 139778}  & K1III:   & $1.072\pm 0.014$ & W1-E2\\
	\object{HD 186815}  & K2III    & $0.713\pm 0.009$ & W1-E2 \\
	\object{HD 206349}  & K1II-III & $0.869\pm 0.011$ & W1-E1, W1-E2\\ 
	\object{HD 206842}  & K1III    & $1.214\pm 0.016$ & W1-E2 \\
	\object{HD 214995}  & K0III:   & $0.947\pm 0.013$ & W1-E1 \\
	\object{HD 216646}  & K0III    & $1.051\pm 0.015$ & W1-E1, W1-E2 \\
	\object{HD 217673}  & K1.5II   & $1.411\pm 0.020$ & W1-E2 \\
	\hline
      \end{tabular}
    \end{center}
    \caption{Calibrators with spectral type, uniform disk angular
      diameter in K band (in milliarcsecond) and baseline
      \citep{Merand2005}. }
    \label{calib1}
\end{table}

The theoretical correction for LD has only a weak influence on the
$p$-factor determination, since that determination is related to a
diameter {\it variation}.  For example, based on our data set, a
general bias of 5\% in the diameters (due to a wrongly estimated limb
darkening) leads to a bias smaller than 1\% in terms of the
\textit{p}-factor.  Differential variations of the LD correction
during the pulsation may also influence the projection factor:
comparison between hydrodynamic and hydrostatic simulations
\citep{Marengo2003} showed negligible variations. An accuracy of 0.2\%
on the angular diameters for a given baseline is required to be
sensitive to dynamical LD effects. This is close to, but still beyond,
the best accuracy that we obtained on the angular diameter with a
single visibility measurement: 0.35\% (median 0.45\%).

Among the various sets of measurements of the radial velocity
$V_\mathrm{rad.}(t)$ available for $\delta$~Cep, we chose measurements
from \citet{Bersier1994} and \citet{Barnes2005}. These works offer the
best phase coverage, especially near the extrema, in order to
accurately estimate the associated photospheric amplitude. In order
not to introduce any bias due to a possible mismatch in the radial
velocity zero-point between the two data sets, we decided to reduce
them separately and then combine the resulting $p$-factor. An
integration over time is required to obtain the photospheric
displacement (see Eq.\ref{bw}). This process is noisy for unequally
spaced data points: the radial velocity profile was smoothly
interpolated using a periodic cubic spline function.


Fitting the inferred photospheric displacement and observed angular
diameter variations, we adjust three parameters: the mean angular
diameter $\overline{\theta}$, a free phase shift $\phi_0$ and the
projection factor $p$ (see Fig.~1).  The mean angular diameter is
found to be $1.475\pm0.004$ mas (milliarcsecond) for both radial
velocity data sets. Assuming a distance of $274 \pm 11$ pc
\citep{Benedict2002}, this leads to a linear radius of $43.3\pm1.7$
solar radii. The fitted phase shift is very small in both cases (of
the order of $0.01$). We used the same parameters \citep{Moffett1985}
to compute the phase from both observation sets and considering that
they were obtained more than ten years apart, this phase shift
corresponds to an uncertainty in the period of approximately five
seconds. We thus consider the phase shift to be reasonably the result
of uncertainty in the ephemeris.

\begin{figure}
  \begin{center}
    \resizebox{\hsize}{!}{\includegraphics{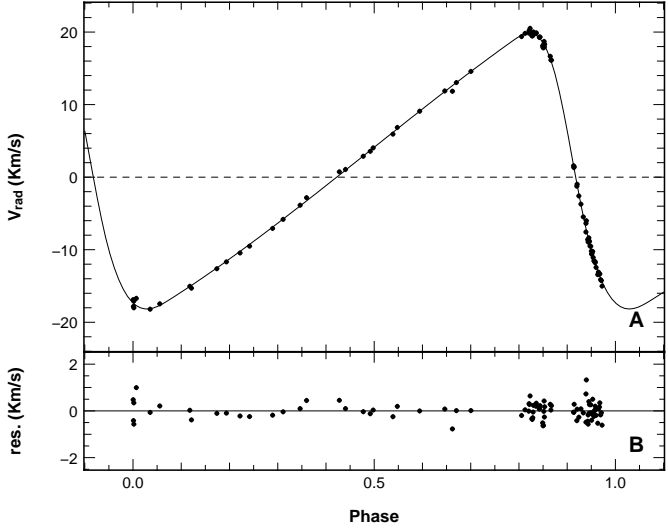}}
    \caption{Radial Velocity smoothed using splines. A. Radial
      velocity data points, as a function of pulsation phase (0-phase
      defined as the maximum of light). This set was extracted using a
      cross-correlation technique \citep{Bersier1994}.  The solid line
      is a 4-knot periodic cubic spline fit. B. Residuals of the
      fit.}
    \label{dc1}
  \end{center}
\end{figure}

\begin{figure}
  \begin{center}
    \resizebox{\hsize}{!}{\includegraphics{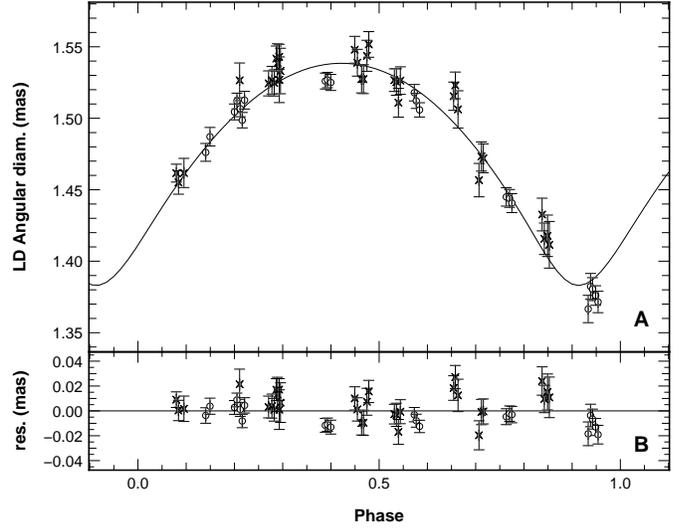}}
    \caption{$p$-factor determination. A. Our angular diameter
      measurements (points). Crosses correspond to the medium baseline
      (E2-W1), while circles correspond to the largest baseline
      (E1-W1). The continuous line is the integration of the 4-knots
      periodic cubic spline fitted to the radial velocities
      (Fig~\ref{dc1}) . Integration parameters:
      $\overline{\theta}=1.475\; \mathrm{mas}$, $p=1.269$ and
      $d=274\;\mathrm{pc}$. B. Residuals of the fit}
    \label{dc2}
  \end{center}
\end{figure}

The two different radial velocity data sets lead to a consolidated
value of $p=1.27\pm0.06$, once again assuming a distance of $274 \pm
11\;\mathrm{pc}$. The final reduced $\chi^2$ is 1.5. The error bars
account for three independent contributions: uncertainties in the
radial velocities, the angular diameters and the distance. The first
was estimated using a bootstrap approach, while the others were
estimated analytically (taking into account calibration correlation
for interferometric errors): for $p$, the detailed error is $p=1.273
\pm 0.007_\mathrm{Vrad.}  \pm 0.020_\mathrm{interf.}  \pm
0.050_\mathrm{dist.}$.  The error is dominated by the distance
contribution (see Table~\ref{table1}).

\begin{table}
    \begin{center}
      \begin{tabular}{rr@{$\;\pm\;$}lc}  
        \hline
	\hline
	$p\pm\sigma_\mathrm{Vrad.}$ & 
	1.269 & 0.008 & ref. (1) \\ 
	 & 1.280 & 0.012 & ref. (2) \\ 
	\hline
        $p\pm \sigma_\mathrm{Vrad.}$ & 1.273 & 0.007 & consolidated \\
        $\sigma_\mathrm{interf.}$ & & 0.020 \\
        $\sigma_\mathrm{dist.}$ & & 0.050 \\
        \hline $p$ & 1.27 & 0.06 \\ 
	\hline  
      \end{tabular}
    \end{center}
    \caption{\label{table1} Best fit results for $p$, with the two
    different radial velocity sets. The third line is a weighted
    average of the two individual measurements. Fourth and fith lines
    are the detailed quadratic contribution to the final error
    bar. Last line gives the final adopted value with the overall
    error bar. References are: (1) \citet{Bersier1994} and (2)
    \citet{Barnes2005}}
\end{table}

\section{Discussion}

Until now, the $p$-factor has been determined using models:
hydrostatic models \citep{Burki1982} produced the generally adopted
value, $p=1.36$.  First attempts were made by \citet{Sabbey1995} to
take into account dynamical effects due to the pulsation. They
concluded that the average value of $p$ should be 5\% larger than in
previous works (1.43 instead of 1.36) and that $p$ is not constant
during the pulsation. Because they increased $p$ by 5\%, they claimed
that distances and diameters have to be larger in the same
proportion. More recently \citet{Nardetto2004} computed $p$
specifically for $\delta$~Cep using dynamical models. Different values
of $p$ were found, whether one measures diameters in the continuum or
in the layer where the specific line is formed. In our case, broad
band stellar interferometry (angular diameters are measured in the
continuum) these authors suggest $p=1.27\pm0.01$. Concerning the
variation of p during the pulsation, they estimate that the error in
terms of distance is of the order of 0.2\%, smaller than what we would
have been able to measure with our interferometric data set. While our
estimate, $p=1.27\pm0.06$, is statistically compatible with this
recent work, marginally with the widely used $p=1.36$, and not
consistent with the former value $p=1.43$ at a 2\,$\sigma$ level. We
note that Gieren et al. (2005) have recently derived an expression of
the $p$-factor as a function of the period that predicts a value of
$1.47 \pm 0.06$ for $\delta$\,Cep. While this value is in agreement
with the modeling by \citet{Sabbey1995}, is is slightly larger than
the present measurement (by 2.4\,$\sigma$). As a remark, Gieren et
al. obtain a distance of $280 \pm 4$\,pc for $\delta$~Cep, that is
slightly larger than Benedict et al.'s (2002) value $274 \pm 11$\,pc
assumed in the present work. Assuming this new distance estimation
with our data would result in a $p$-factor of $1.30 \pm 0.06$,
bringing the agreement to 2\,$\sigma$ only.

Our geometrical determination of the $p$-factor, $p=1.27\pm0.06$, using
the IBW method is currently limited by the error bar on the
parallax \citep{Benedict2002}. Conversely, assuming a perfectly known
$p$-factor, the uncertainty of the stellar distance determined using the
same method would have been only 1.5\%, two-times better than
the best geometrical parallax currently available. The value we
determined for $p$ is statistically compatible with the value
generally adopted to calibrate the Cepheid P-L relation in most
recent works. It is expected that the distance to approximatively 30
Cepheids will be determined interferometrically in the near future
using particularly the CHARA Array and the VLT
Interferometer \citep{VLTI2005}. In order not to limit the final
accuracy on the derived distances, theoretical $p$-factor studies
using realistic hydrodynamical codes is necessary.  With a better
understanding of the detailed dynamics of the Cepheid atmospheres, we
will be in a position to exclude a $p$-factor bias on the calibration
of the P-L relation, at a few percent level.

\begin{acknowledgements}
We thank P.J.~Goldfinger for her assistance during the
observations. The CHARA Array was constructed with funding from
Georgia State University, the National Science Foundation, the
W. M. Keck Foundation, and the David and Lucile Packard
Foundation. The CHARA Array is operated by Georgia State University
with support from the College of Arts and Sciences, from the Research
Program Enhancement Fund administered by the Vice President for
Research, and from the National Science Foundation under NSF Grant
AST~0307562.
\end{acknowledgements}

\bibliographystyle{aa}
\bibliography{He061.bib}
\clearpage

\onecolumn
\begin{landscape}
\begin{longtable}{lcccccrccrcc}

        \hline
	\hline
	JD-JD$_0$ &$\phi$ & U(m) & V(m) & $V^2$ (\%) & $\theta_\mathrm{LD}$
	(mas) & $HD_a$ & $\alpha$ & $\sigma_{V^2_a}$& $HD_b$ &
	$\beta$ & $\sigma_{V^2_b}$ \\
	
	\hline
	16.3844 & 0.289 & -246.23 &-13.93  & $12.09\pm0.58$ & $1.539\pm0.014$ & 206842 & 0.232 & 0.0106 & 217673 & 0.313 & 0.0096 \\ 
	16.4051 & 0.293 & -245.91 &-41.24  & $11.94\pm0.69$ & $1.526\pm0.016$ & 217673 & 0.354 & 0.0096 & 217673 & 0.322 & 0.0095 \\ 
	17.3801 & 0.475 & -246.08 &-11.83  & $12.79\pm0.47$ & $1.524\pm0.011$ & 217673 & 0.096 & 0.0102 & 216646 & 0.270 & 0.0114 \\ 
	17.4005 & 0.478 & -246.11 &-38.71  & $11.89\pm0.37$ & $1.529\pm0.009$ & 216646 & 0.154 & 0.0114 & 216646 & 0.152 & 0.0114 \\ 
	18.3443 & 0.654 & -237.44 & 31.26  & $16.33\pm0.48$ & $1.489\pm0.010$ & 216646 & 0.188 & 0.0111 & 216646 & 0.189 & 0.0112 \\ 
	18.3630 & 0.658 & -243.62 & 7.08   & $14.64\pm0.43$ & $1.499\pm0.009$ & 216646 & 0.173 & 0.0112 & 216646 & 0.183 & 0.0114 \\ 
	18.3935 & 0.663 & -246.44 &-33.07  & $13.63\pm0.59$ & $1.491\pm0.013$ & 217673 & 0.373 & 0.0096 & 216646 & 0.177 & 0.0114 \\ 
	19.3289 & 0.838 & -231.37 & 47.22  & $21.53\pm0.63$ & $1.407\pm0.011$ &      0 & - & - & 216646 & 0.474 & 0.0110 \\ 
	19.3536 & 0.842 & -241.79 & 15.84  & $20.50\pm0.61$ & $1.390\pm0.011$ & 216646 & 0.209 & 0.0110 & 216646 & 0.262 & 0.0112 \\ 
	19.3889 & 0.849 & -246.53 &-30.66  & $17.95\pm0.77$ & $1.403\pm0.015$ & 217673 & 0.417 & 0.0098 & 216646 & 0.250 & 0.0114 \\ 
	19.4093 & 0.853 & -243.71 &-57.56  & $17.59\pm0.86$ & $1.399\pm0.016$ & 216646 & 0.214 & 0.0114 & 217673 & 0.526 & 0.0095 \\ 
	21.3301 & 0.211 & -234.72 & 38.94  & $17.02\pm0.57$ & $1.484\pm0.012$ & 216646 & 0.296 & 0.0095 &      0 & - & - \\ 
	28.4176 & 0.531 & -230.78 &-99.51  & $11.94\pm0.36$ & $1.514\pm0.008$ & 206349 & 0.111 & 0.0089 & 216646 & 0.153 & 0.0114 \\ 
	28.4406 & 0.536 & -215.78 &-127.18 & $12.38\pm0.41$ & $1.509\pm0.010$ & 216646 & 0.272 & 0.0114 & 206349 & 0.030 & 0.0083 \\ 
	28.4630 & 0.540 & -196.83 &-152.03 & $12.49\pm0.47$ & $1.517\pm0.011$ & 216646 & 0.171 & 0.0114 & 206349 & 0.099 & 0.0083 \\ 
	28.4848 & 0.544 & -174.70 &-173.74 & $12.24\pm0.48$ & $1.537\pm0.011$ & 216646 & 0.060 & 0.0114 & 206349 & 0.169 & 0.0083 \\ 
	29.3593 & 0.707 & -246.59 &-27.63  & $15.93\pm0.57$ & $1.445\pm0.012$ & 206842 & 0.318 & 0.0106 & 216646 & 0.186 & 0.0114 \\ 
	29.3863 & 0.712 & -242.60 &-63.16  & $14.92\pm0.49$ & $1.451\pm0.010$ & 216646 & 0.161 & 0.0114 & 216646 & 0.224 & 0.0114 \\ 
	29.4074 & 0.716 & -234.58 &-90.27  & $14.80\pm0.48$ & $1.450\pm0.010$ & 216646 & 0.385 & 0.0114 &      0 & - & - \\ 
	31.3590 & 0.080 & -246.38 &-34.41  & $15.37\pm0.38$ & $1.453\pm0.008$ & 186815 & 0.099 & 0.0071 & 206349 & 0.165 & 0.0090 \\ 
	31.3828 & 0.084 & -242.03 &-65.75  & $15.39\pm0.40$ & $1.441\pm0.008$ & 206349 & 0.126 & 0.0090 & 216646 & 0.226 & 0.0114 \\ 
	31.4433 & 0.095 & -207.08 &-139.54 & $15.96\pm0.51$ & $1.435\pm0.010$ & 216646 & 0.415 & 0.0114 &      0 & - & - \\ 
	32.3850 & 0.271 & -240.48 &-72.01  & $12.49\pm0.38$ & $1.503\pm0.009$ & 138852 & 0.050 & 0.0094 & 216646 & 0.260 & 0.0114 \\ 
	32.4220 & 0.278 & -221.30 &-118.17 & $12.68\pm0.44$ & $1.500\pm0.010$ & 216646 & 0.139 & 0.0114 & 216646 & 0.183 & 0.0114 \\ 
	32.4470 & 0.282 & -201.55 &-146.52 & $13.09\pm0.42$ & $1.501\pm0.009$ & 216646 & 0.168 & 0.0114 & 216646 & 0.152 & 0.0112 \\ 
	32.4710 & 0.287 & -177.75 &-171.09 & $12.86\pm0.42$ & $1.520\pm0.010$ & 216646 & 0.173 & 0.0112 & 216646 & 0.125 & 0.0110 \\ 
	32.5025 & 0.293 & -140.59 &-198.07 & $13.78\pm0.43$ & $1.523\pm0.010$ & 216646 & 0.121 & 0.0110 & 216646 & 0.171 & 0.0107 \\ 
	33.3435 & 0.449 & -246.55 &-21.14  & $12.40\pm0.41$ & $1.527\pm0.010$ & 139778 & 0.062 & 0.0102 & 216646 & 0.245 & 0.0114 \\
	33.3723 & 0.455 & -243.42 &-59.13  & $11.72\pm0.47$ & $1.525\pm0.011$ & 216646 & 0.183 & 0.0114 & 206349 & 0.084 & 0.0089 \\ 
	33.4189 & 0.463 & -221.54 &-117.75 & $12.58\pm0.45$ & $1.502\pm0.010$ & 216646 & 0.151 & 0.0114 & 216646 & 0.169 & 0.0114 \\ 
	33.4404 & 0.467 & -204.89 &-142.38 & $12.56\pm0.41$ & $1.511\pm0.009$ & 216646 & 0.315 & 0.0114 &      0 & - & - \\ 
	80.3020 & 0.200 & 253.48 & 183.15  & $2.41\pm0.12$  & $1.491\pm0.005$ & 185395 & 0.008 & 0.0139 & 216646 & 0.086 & 0.0112 \\ 
	80.3295 & 0.205 & 220.84 & 218.17  & $2.46\pm0.12$  & $1.500\pm0.006$ & 216646 & 0.078 & 0.0112 &   2952 & 0.022 & 0.0112 \\ 
	80.3667 & 0.212 & 166.28 & 257.01  & $2.85\pm0.13$  & $1.502\pm0.006$ &   2952 & 0.046 & 0.0112 &   2952 & 0.049 & 0.0112 \\ 
	80.3888 & 0.216 & 129.38 & 274.56  & $3.20\pm0.13$  & $1.498\pm0.006$ &   2952 & 0.050 & 0.0112 &   2952 & 0.056 & 0.0112 \\ 
	80.4145 & 0.221 & 83.29 & 289.27   & $3.19\pm0.15$  & $1.511\pm0.006$ &   2952 & 0.088 & 0.0112 &  37128 & 0.009 & 0.0409 \\ 
	81.3127 & 0.388 & 238.39 & 200.90  & $2.12\pm0.11$  & $1.511\pm0.006$ & 216646 & 0.040 & 0.0110 & 216646 & 0.051 & 0.0112 \\ 
	81.3371 & 0.393 & 206.72 & 230.09  & $2.28\pm0.12$  & $1.514\pm0.006$ & 216646 & 0.083 & 0.0112 &   2952 & 0.009 & 0.0112 \\ 
	81.3739 & 0.400 & 149.99 & 265.47  & $2.61\pm0.13$  & $1.519\pm0.006$ & 216646 & 0.052 & 0.0112 &   2952 & 0.044 & 0.0112 \\ 
	82.3031 & 0.573 & 246.42 & 191.85  & $2.32\pm0.13$  & $1.498\pm0.006$ & 216646 & 0.053 & 0.0110 & 216646 & 0.049 & 0.0112 \\ 
	82.3246 & 0.577 & 220.02 & 218.91  & $2.45\pm0.12$  & $1.501\pm0.006$ & 216646 & 0.057 & 0.0112 &   2952 & 0.037 & 0.0112 \\ 
	82.3611 & 0.584 & 166.58 & 256.84  & $2.78\pm0.11$  & $1.504\pm0.005$ &   2952 & 0.033 & 0.0112 &   2952 & 0.060 & 0.0112 \\ 
	83.3260 & 0.764 & 214.60 & 223.63  & $3.73\pm0.17$  & $1.445\pm0.006$ & 214995 & 0.026 & 0.0097 &   2952 & 0.107 & 0.0112 \\ 
	83.3625 & 0.770 & 159.85 & 260.49  & $4.29\pm0.18$  & $1.444\pm0.006$ &   2952 & 0.063 & 0.0112 &   2952 & 0.080 & 0.0112 \\ 
	83.3878 & 0.775 & 116.78 & 279.31  & $4.75\pm0.20$  & $1.440\pm0.007$ &   2952 & 0.080 & 0.0112 &   2952 & 0.076 & 0.0112 \\ 
	84.2374 & 0.933 & 294.80 & 103.39  & $6.53\pm0.36$  & $1.342\pm0.010$ & 216646 & 0.171 & 0.0109 & 216646 & 0.139 & 0.0109 \\ 
	84.2635 & 0.938 & 278.68 & 143.65  & $5.77\pm0.32$  & $1.359\pm0.009$ & 216646 & 0.136 & 0.0109 & 216646 & 0.138 & 0.0110 \\ 
	84.2855 & 0.942 & 259.31 & 175.36  & $5.64\pm0.28$  & $1.365\pm0.008$ & 216646 & 0.178 & 0.0110 &   2952 & 0.061 & 0.0112 \\ 
	84.3201 & 0.949 & 218.81 & 219.98  & $5.68\pm0.24$  & $1.376\pm0.007$ &   2952 & 0.101 & 0.0112 &   2952 & 0.088 & 0.0112 \\ 
	84.3468 & 0.954 & 180.34 & 248.69  & $6.29\pm0.27$  & $1.371\pm0.008$ &   2952 & 0.118 & 0.0112 &   2952 & 0.093 & 0.0112 \\ 
	85.3490 & 0.140 & 172.71 & 253.33  & $3.36\pm0.16$  & $1.476\pm0.006$ & 176598 & 0.013 & 0.0105 &   2952 & 0.097 & 0.0112 \\ 
	85.3962 & 0.149 & 91.77 & 287.10   & $3.70\pm0.17$  & $1.487\pm0.006$ &   2952 & 0.066 & 0.0112 &   2952 & 0.053 & 0.0112 \\ 
	\hline
	
	\hline

    \caption{Individal measurements. colums are (1) date of
    observation, JD$_0$=$2\; 453\; 200.5$ (2) phase (3,4) u-v coordinate in
    meter (5) squared visibility and error (6) corresponding limb
    darkened disk diameter in mas (7,10) HD number of calibrators,
    prior and after the given data point respectivaly, 0 means that
    there was no calibrator (8,9,11,12) quantities for computing the
    correlation matrix \citep{Perrin2003}: $\sigma_{V^2}$ are
    errors on the estimated visibility of the calibrators. \tt NOTE: this
    table will be archived electronically }
    \label{data1}
\end{longtable}
\end{landscape}


\end{document}